\documentclass[twocolumn,prl,aps,psfig,showpacs,preprintnumbers,superscriptaddress]{revtex4}
\usepackage{graphicx}
\usepackage{epstopdf}
\usepackage{float}
\usepackage{color}
\usepackage{amsmath}
\usepackage{multirow}

\newcommand{\CaK}{CaKFe$_4$As$_4$}

\begin{document}

\title{Magnetic fluctuations and superconducting properties of CaKFe$_4$As$_4$ studied by $^{75}$As NMR}

\author{J.~Cui}
\affiliation{Ames Laboratory, U.S. DOE, Ames, IA 50011, USA}
\affiliation{Department of Chemistry, Iowa State University, Ames, Iowa 50011, USA}
\author{Q.-P.~Ding}
\affiliation{Ames Laboratory, U.S. DOE, Ames, IA 50011, USA}
\author{W.~R.~Meier}
\affiliation{Ames Laboratory, U.S. DOE, Ames, IA 50011, USA}
\affiliation{Department of Physics and Astronomy, Iowa State University, Ames, Iowa 50011, USA}
\author{A.~E.~B\"{o}hmer}
\affiliation{Ames Laboratory, U.S. DOE, Ames, IA 50011, USA}
 \author{T.~Kong*}
\affiliation{Ames Laboratory, U.S. DOE, Ames, IA 50011, USA}
\affiliation{Department of Physics and Astronomy, Iowa State University, Ames, Iowa 50011, USA}
\author{V. Borisov}
\affiliation{Institute of Theoretical Physics, Goethe University Frankfurt am Main, D-60438 Frankfurt am Main, Germany}
\author{Y.~Lee}
\affiliation{Ames Laboratory, U.S. DOE, Ames, IA 50011, USA}
\author{S.~L.~Bud'ko}
\affiliation{Ames Laboratory, U.S. DOE, Ames, IA 50011, USA}
\affiliation{Department of Physics and Astronomy, Iowa State University, Ames, Iowa 50011, USA}
\author{R. Valent\'i}
\affiliation{Institute of Theoretical Physics, Goethe University Frankfurt am Main, D-60438 Frankfurt am Main, Germany}
\author{P.~C.~Canfield}
\affiliation{Ames Laboratory, U.S. DOE, Ames, IA 50011, USA}
\affiliation{Department of Physics and Astronomy, Iowa State University, Ames, Iowa 50011, USA}
\author{Y.~Furukawa,}
\affiliation{Ames Laboratory, U.S. DOE, Ames, IA 50011, USA}
\affiliation{Department of Physics and Astronomy, Iowa State University, Ames, Iowa 50011, USA}

\date{\today}

\begin{abstract}

       We report $^{75}$As nuclear magnetic resonance (NMR) studies on a new iron-based superconductor CaKFe$_4$As$_4$ with $T_{\rm c}$ = 35 K. 
       $^{75}$As NMR spectra show two distinct lines corresponding to the As(1) and As(2) sites close to the K and Ca layers, respectively, revealing   that K and Ca layers are well ordered without site inversions. 
     We found that nuclear quadrupole frequencies $\nu_{\rm Q}$ of the As(1) and As(2) sites show an opposite temperature ($T$) dependence.
       Nearly $T$ independent behavior of the Knight shifts $K$ are observed in the normal state,  and a sudden decrease in $K$ in the superconducting (SC) state suggests spin-singlet Cooper pairs. 
    $^{75}$As spin-lattice relaxation rates 1/$T_1$ show a power law $T$ dependence with different exponents for the two As sites.
       The isotropic antiferromagnetic spin fluctuations characterized by the wavevector ${\bf q}$ = ($\pi$, 0) or (0, $\pi$) in the single-iron Brillouin zone notation 
are revealed by 1/$T_1T$ and $K$ measurements. 
Such magnetic fluctuations are necessary to explain the observed temperature dependence of the $^{75}$As quadrupole frequencies, as evidenced by our first-principles calculations. 
         In the SC state, 1/$T_1$ shows a rapid decrease below $T_{\rm c}$ without a Hebel-Slichter peak and decreases exponentially  at low $T$, consistent with an $s^{\pm}$ nodeless two-gap superconductor. 
  
\end{abstract}

\maketitle

  \section{I. Introduction} 
       Following the discovery of unconventional superconductivity (SC) in the so-called 1111 and 122 systems exemplified by LaFeAsO \cite{Kamihara2008} and BaFe$_2$As$_2$ \cite{Rotter2008}, respectively, many pnictides have been investigated in the search for unconventional superconductivity and other novel properties \cite{Johnston2010, Canfield2010, Ni2008,Chu2009,Stewart2011}.
              Quite recently, a new Fe-based SC  CaKFe$_4$As$_4$ (in short, CaK1144) with a transition temperature of $T_{\rm c}$ $\sim$ 33.1 K has been discovered \cite{Iyo2016}. 
              In CaK1144, Ca$^{2+}$ and K$^{1+}$ occupy different atomic positions in its crystal structure due to the different ionic radii, forming alternating Ca and K planes along the crystallographic $c$  axis separated by FeAs layers \cite{Iyo2016}.
        This is contrary to the doped 122 system such as Ba$_{1-x}$K$_x$Fe$_2$As$_2$ where alkaline-earth metal and alkaline ions occupy randomly the same atomic position.
        The ordering of the K and Ca ions changes the space group from $I4/mmm$  system to $P4/mmm$.
        Consequently, as shown in the inset of  Fig.\ \ref{fig:As-spectrum}(a), there are two different As sites: As(1) and As(2) sites close to the K and Ca layers, respectively.

        Soon after the discovery of CaK1144, thermodynamic and transport measurements on single crystals revealed that CaK1144 is an ordered stoichiometric superconductor with $T_{\rm c}$ $\sim$ 35 K and a very high upper critical field of 92 T \cite{Meier2016} with no other phase transition from 1.8 K to room temperature, and shows similar physical properties to the optimally doped (Ba$_{1-x}$K$_x$)Fe$_2$As$_2$.  
       Muon spin rotation and relaxation ($\mu$SR) measurements show two nodeless gaps in CaK1144 with gap sizes of 2.5 and 8.6 meV \cite{uSR2017}, consistent with tunnel-diode resonator (TDR) and scanning tunneling microscope (STM) measurements \cite{Cho2017}.  
       The multiple superconducting gaps in CaK1144 are also revealed by high resolution angle resolved photoemission spectroscopy (ARPES) measurements and density functional theory (DFT) calculations \cite{ARPES2016}.
       However, studies of magnetic fluctuations in CaK1144 have not been carried out yet.  
       Since the magnetic fluctuations are considered to be one of the keys in driving SC in iron pnictide superconductors \cite{Johnston2010, Ishida2009,Paglione2010, PaulPRL,JeanPRB}, it is important to characterize the dynamical properties of the Fe moments.
  
     Nuclear magnetic resonance (NMR) is a microscopic probe suitable for investigating static spin susceptibility and  low-energy spin excitations for pnictide superconductors \cite{Johnston2010,Ishida2009}.
     The temperature $T$ dependence of the nuclear spin-lattice relaxation rate (1/$T_1$)  is known to reflect the wave vector $q$-summed dynamical susceptibility. 
     On the other hand, NMR spectrum measurements, in particular the Knight shift $K$, give us information on static magnetic susceptibility $\chi$. 
     Thus from the $T$ dependence of 1/$T_1T$ and $K$, one can obtain valuable insights about magnetic fluctuations in materials.  
     Furthermore, 1/$T_1$ measurements in the SC state provide important information in understanding the gap structure in SCs.        

       In this paper, we carried out $^{75}$As NMR measurements to investigate the electronic and magnetic properties of single crystalline CaKFe$_4$As$_4$ with $T_{\rm c}$ = 35~K. 
       We observed two $^{75}$As NMR signals corresponding to the two different As sites in the crystal structure, showing the ordering of Ca and K ions on each layer.  
      The antiferromagnetic (AFM) spin fluctuations are clearly evidenced by $^{75}$As spin-lattice relaxation rate 1/$T_1$ and Knight shift $K$ measurements.
	  \textit{Ab initio} density functional theory calculations of the
 $^{75}$As nuclear quadrupole frequency ($\nu_{\rm Q}$) confirm the importance of such fluctuations for the
 description of the observed temperature evolution of $\nu_Q$.
         In the superconducting state, 1/$T_1$ shows a rapid decrease below $T_{\rm c}$ without showing a Hebel-Slichter coherence peak and decreases exponentially at low temperatures, which is explained by a two-nodeless-gap model, consistent with previous reports.

  \section{II. Experimental methods}

         The single crystals of  CaKFe$_4$As$_4$  for NMR measurements  were grown by high temperature solution growth out of FeAs flux \cite{Meier2016}  and were extensively characterized by thermodynamic and transport measurements \cite{Meier2016}.
    NMR measurements were carried out on $^{75}$As  (\textit{I} = 3/2, $\gamma/2\pi$ = 7.2919 MHz/T, $Q$ =  0.29 Barns)  by using a lab-built, phase-coherent, spin-echo pulse spectrometer.  
   The $^{75}$As-NMR spectra were obtained by sweeping the magnetic field at a fixed frequency $f$ = 53 MHz.
    The magnetic field was applied parallel to either the crystalline $c$ axis or the $ab$ plane. 
    The $^{75}$As 1/$T_{\rm 1}$ was measured with a recovery method using a single $\pi$/2 saturation pulse.
 In the normal state above $T_{\rm c}$, the $1/T_1$ at each $T$ was determined by fitting the nuclear magnetization $M$ versus time $t$  using the exponential function $1-M(t)/M(\infty) = 0.1 e^ {-t/T_{1}} +0.9e^ {-6t/T_{1}}$ for $^{75}$As NMR, where $M(t)$ and $M(\infty)$ are the nuclear magnetization at time $t$ after the saturation and the equilibrium nuclear magnetization at $t$ $\rightarrow$ $\infty$, respectively. 
   In the SC state, the function could not reproduce the $M$ versus  $t$ data. 
   Then we fitted the data with the two $T_1$ component function 
$1-M(t)/M(\infty) = M_{\rm S} [0.1 e^ {-t/T_{1 \rm S}} +0.9e^ {-6t/T_{1 \rm S}}] + M_{\rm L} [0.1 e^ {-t/T_{1 \rm L}} +0.9e^ {-6t/T_{1 \rm L}}] $ where $M_{\rm S}+M_{\rm L} = 1$. 
    The $M_{\rm S}$  and $M_{\rm L}$ correspond to the fraction for the short relaxation time $T_{\rm 1S}$ and the long relaxation time $T_{\rm 1L}$, respectively. 
  The $M_{\rm L}$ was estimated to be $\sim$ 0.5 in a temperature range of 4.3 - 25 K and then starts to increase slowly up to 1 at $T_{\rm c}$. The $T_{\rm 1S}$ and $T_{\rm 1L}$ could be attributed to $T_1$ for the As(2) and As(1) sites, respectively, in the SC state.


 \section{III. Results and discussion}
\subsection{A. $^{75}$As NMR spectrum}

\begin{figure}[tb]
\includegraphics[width=8.5 cm]{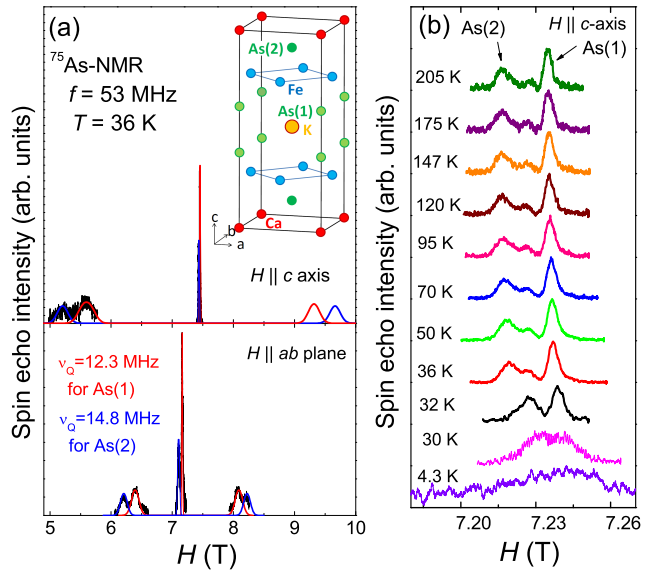}
\caption{(a) Field-swept $^{75}$As-NMR spectra at $f=53$ MHz for magnetic fields $H\parallel ab$ plane and $H\parallel c$ axis at $T$ = 36 K. 
The red and blue lines are simulated spectra for the As(1) and As(2) sites, respectively. Inset shows the crystal structure of CaKFe$_4$As$_4$.
(b) Temperature dependence of the central transition lines of $^{75}$As-NMR spectra for the As(1) and As(2) sites for $H\parallel c$ axis.
}
\label{fig:As-spectrum}
\end{figure}

\begin{figure}[tb]
\centering
\includegraphics[width=8.5cm]{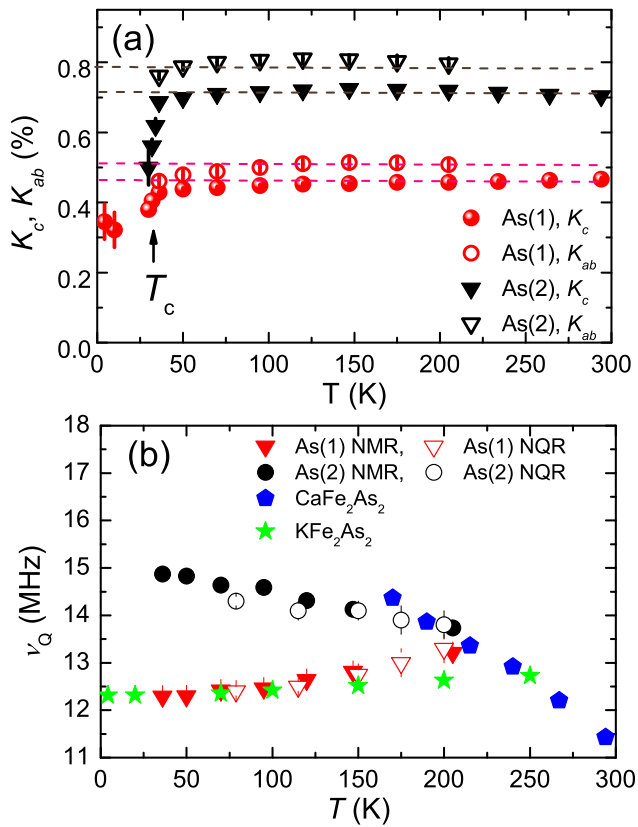}
\caption{(a) Temperature dependences of the $^{75}$As-NMR shifts $K_c$ and $K_{ab}$ for the As(1) and As(2) sites.  
(b) Temperature dependences of quadrupole frequency $\nu_{\rm Q}$ for the As(1) and the As(2) sites, estimated from NMR and NQR spectra measurements. The $\nu_{\rm Q}$s above 205 K were not determined due to poor signal intensity. 
  $^{75}$As-$\nu_{\rm Q}$ data for KFe$_2$As$_2$ \cite{Wiecki_unpublish} and CaFe$_2$As$_2$ \cite{Cui2015} are also plotted for comparison. }
\label{fig:K}
\end{figure}

     Figure\ \ref{fig:As-spectrum}(a)  shows typical field-swept $^{75}$As-NMR spectra of the CaKFe$_4$As$_4$ single crystal at  $T$ = 36~K   for the two magnetic field directions: $H$ $\parallel$ $c$ axis  and $H$ $\parallel$ $ab$ plane. 
   The typical spectrum for a nucleus with spin $I=3/2$ with Zeeman and quadrupolar interactions can be described by a nuclear spin Hamiltonian ${\cal{H}}=-\gamma\hbar(1+K)HI_z+\tfrac{h\nu_{\rm Q}}{6}[3I_z^2-I(I+1)+\frac{1}{2}\eta(I^{2}_{+} + I^{2}_{-})]$, where $K$ is the Knight shift, $h$ is Planck's constant, and $\hbar = h/2\pi$. 
   The nuclear  quadrupole frequency for $I=3/2$ nuclei is given by $\nu_{\rm Q} = eQV_{\rm ZZ}/2h\sqrt{1+\eta^2/3}$, where $Q$ is the nuclear quadrupole moment, $V_{\rm ZZ}$ is the electric field gradient (EFG) along $z$ at the As site and $\eta = (V_{\rm YY} - V_{\rm XX})/V_{\rm ZZ} $ is the in-plane asymmetry of the EFG tensor.
    For $I=3/2$ nuclei, when the Zeeman interaction is much greater than the quadruple interaction, this Hamiltonian produces a spectrum with a sharp central transition line flanked by single satellite peak on each side for each equivalent As site. 
    As shown in the crystal structure [inset of Fig.\ \ref{fig:As-spectrum}(a)], CaKFe$_4$As$_4$ has two inequivalent As sites: As(1) and As(2) close to the K and Ca layers, respectively.
    As seen in  Fig.~\ \ref{fig:As-spectrum}(a), we indeed observed two sets of $I$ =3/2 quadrupole split lines corresponding to the two As sites for each $H$ direction.
    The observed $^{75}$As NMR spectra are well reproduced by simulated spectra [the red and blue lines in Fig.\ \ref{fig:As-spectrum}(a)] from the above simple Hamiltonian where the Zeeman interaction is greater than the quadrupole interaction. 
     Since the distance between the As(1) and Fe ions is 2.395(2) \AA~  which is slightly greater than 2.387(2)~\AA~  for the As(2) sites at 300 K \cite{ARPES2016}, the As(1) site is expected to be less strongly coupled to the Fe ions than the As(2) sites.
    Thus we assign the lower field central peak with a greater Knight shift $K$ to the As(2) site and higher field central peak with a smaller $K$ to the As(1) sites, respectively [see, Fig. \ \ref{fig:As-spectrum}(b)].
    The well separated As(1) and As(2) lines indicate that K and Ca ions are not randomly  distributed but are well ordered making the Ca and K layers without a significant  site inversion. 
     The full-width at half-maximum (FWHM) of each satellite line is estimated to be 1.35 kOe and 1.40 kOe for the As(1) and As(2) sites at 36 K and $H \parallel ab$ plane, respectively.
   Since the FWHM of the satellite line is proportional to the distribution of $\nu_{\rm Q}$ and thus $V_{ZZ}$, the results indicate no large difference in inhomogeneity for the local As environments, again consistent with the ordering of the Ca and K ions for each plane.
    It is noted that these values are greater than $\sim$ 0.75 kOe for the satellite line of $^{75}$As NMR at 200 K and $H \parallel  ab$  plane in CaFe$_2$As$_2$ which exhibits a stripe-type antiferromagnetic order below $T_{\rm N}$ $\sim$ 170 K \cite{Furukawa2014} and are also greater than $\sim$ 0.4 kOe at 10 K in the superconductor KFe$_2$As$_2$ with $T_{\rm c}$  = 3.5 K \cite{Wiecki_unpublish}.
    On the other hand, these values are smaller than the K-doped BaFe$_2$As$_2$ case, for example, $\sim$ 2.5 kOe at 50 K in  Ba$_{0.42}$K$_{0.58}$Fe$_2$As$_2$ \cite{Hirano2012}.

     The $T$ dependence of the central transition lines of the As(1) and As(2) sites under $H$ parallel to the $c$ axis is shown in 
Fig.~\ \ref{fig:As-spectrum}(b) where no obvious change in the spectra has been observed down to $T_{\rm c}$ = 33 K which is slightly reduced from the zero-field transition temperature due to the application of $H$ $\sim$ 7.23~T.  
     The FWHM of the central line is nearly independent of $T$ with $\sim$46 and $\sim$85 Oe for the As(1) and As(2) sites, respectively, in the normal state, indicating no magnetic and structural phase transitions above $T_{\rm c}$. 
     The difference in FWHM originates from the distribution of the hyperfine field at the As sites. 
    The small hump between the As(1) and As(2) lines is not intrinsic, probably due to unknown impurity, since no angle and temperature dependences are observed.
    Below $T_{\rm c}$, the intrinsic two peaks shift to higher magnetic field and, at the same time, the lines became broad due to the SC transition.
    Even at 30 K, just 3 K below $T_{\rm c}$, the two lines were smeared out and no clear separation of the two lines is observed as seen in Fig. \ \ref{fig:As-spectrum}(b).
    Despite a very low signal intensity, we were able to measure the spectrum at 4.2 K  which is very broad but still allows to determine the peak position. 
     
    Figure\ \ref{fig:K}(a) shows the $T$ dependence of NMR shifts $K_{ab}$ ($H\parallel ab$ plane)  and $K_c$  ($H\parallel c$ axis) for the two As sites. 
    For both As sites, $K$'s are nearly independent of $T$ above $\sim$130~K and decrease very slightly down to $T_{\rm c}$ upon cooling.
    Similar weak $T$ dependence of $K$ is observed in (Ba$_{1-x}$K$_x$)Fe$_2$As$_2$ \cite{Hirano2012}.
    Below  $T_{\rm c}$, $K_c$ for both  the As sites show a sudden drop, indicating spin-singlet Cooper pairing. 


\subsection{B. $^{75}$As quadrupole frequency}

      Figure \ref{fig:K}(b) shows the $T$ dependence of  $\nu_{\rm Q}$ for the two As sites estimated from the fit of the NMR spectra, together with the $^{75}$As-$\nu_{\rm Q}$ data in KFe$_2$As$_2$ and CaFe$_2$As$_2$.  
     The  $\nu_{\rm Q}$s for the As(1) and the As(2) sites are also obtained by nuclear quadrupole resonance (NQR) measurements under zero magnetic field.  
       For the As(2) site, with decreasing $T$, $\nu_{\rm Q}$ increases from 13.8~MHz at 205 K to 14.8~MHz at 36 K, while for the As(1) site an opposite trend is observed and   $\nu_{\rm Q}$ decreases from 13.2 MHz at 205~K to 12.2 MHz at 36~K with decreasing $T$. 
      It is interesting to note that the $T$ dependences of $\nu_{\rm Q}$ of As(1) and As(2) are similar to those of $\nu_{\rm Q}$ of the As site in KFe$_2$As$_2$ \cite{Wiecki_unpublish} and CaFe$_2$As$_2$ \cite{Cui2015}, respectively, as shown in the figure.
     These results are also consistent with our site assignment.

In order to understand the peculiar temperature dependence of  $\nu_{\rm Q}$ for As(1) and As(2), we have performed a density functional study of the electric-field gradients in \CaK{} under different assumptions concerning the structural and spin degrees of freedom.

    By making use of the experimental lattice parameters measured between 6 K and 300 K \cite{Meier2016}, two types of structures were created due to the present lack of knowledge of the internal positions of As and Fe; one type was obtained by fixing the $z_\text{Fe}$  and $z_\text{As}$ to their values known for the 300 K structure ($z_\text{Fe} = 0.2319$; As:  $z_\text{Ca} = 0.1235$ and $z_\text{K} = 0.3415$) and the volume adapted to the experimental  data at each temperature. This procedure is equivalent to a uniform stretching of the crystal 
when going to higher temperatures. The second type of crystal structures was obtained by fixing the volume given at each temperature and relaxing the internal atomic positions. 
    Magnetic fluctuations are taken into account -in a first approximation- by imposing different types of {\it frozen} magnetic  order. 
    This approach has been successfully applied to the recently discovered half-collapse transition in CaKFe$_4$As$_4$ where a specific type of spin configuration reproduces the structural behavior under external pressure \cite{Kaluarachchi2017}.
    In the present study, we analyze the effect of such (fluctuating) spin configurations on the electric field gradients (EFG) $V_{\rm ZZ}$ on As sites. 
    The latter  were calculated fully relativistically 
in the generalized-gradient approximation (GGA) using the projector-augmented wave pseudopotential 
method of the Vienna Ab initio Simulation Package 
(VASP)\,\cite{Kresse1993PRB,Kresse1996PRB,Kresse1996,Bloechl1994,Kresse1999}. 
The integration over the irreducible Brillouin zone was realized on the $\Gamma$-centered 
$(10\times 10\times 10)$ k-mesh. The $V_{\rm ZZ}$ values are converted to the quadrupole frequencies 
using the relation $ \nu_{\rm Q} = eQV_{\rm ZZ}/2h\sqrt{1 + \eta^2/3}$, discussed already in Sec.~III\,A. 
For clarity, we first consider here in detail the spin-vortex hedgehog phase \cite{Meier2017} and compare 
its behavior to the purely non-magnetic case. This type of order has one of the lowest energies among 
the configurations preserving the $C_4$ symmetry and has been determined already to play a role for 
pressure-induced structural collapse transitions in \CaK \cite{Kaluarachchi2017}, as well as
in Co and Ni-doped CaK1144 where a hedgehog long range order has been identified \cite{Meier2017}.

\begin{table*}[tb]
  \caption{Temperature dependence of Fe magnetic moments for the \CaK{} hedgehog structures depicted in Fig.~\ref{f:NMR_vs_temperature}(a,i) with optimized Fe and As positions is shown in comparison with the Fe moments for the case when the internal positions are fixed at their room-temperature values ($z_\text{Fe} = 0.2319$; As: $z_\text{Ca} = 0.1235$ and $z_\text{K} = 0.3415$). The measured lattice parameters $a$ and $c$ from Ref.~[\onlinecite{Meier2016}] are given at each temperature.}
\vspace{5pt}
\centering
\setlength{\tabcolsep}{12pt}
\renewcommand{\arraystretch}{1.15}
   \begin{tabular}{c|c|ccc|cc}\hline\hline
          & fixed $z_\text{As}$ & \multicolumn{3}{c|}{optimized $z_\text{As}$} & \multirow{2}{*}{a (\AA)} & \multirow{2}{*}{c (\AA)}  \\ \cline{1-5}
      T (K)  &  $m_\text{Fe} \,(\mu_B)$  &  $m_\text{Fe} \,(\mu_B)$  &   $z_\text{Ca}$  &  $z_\text{K}$   \\ \hline
        6    &     1.51    &    1.32  &  0.12318 &  0.33286 &  3.861  &  12.675  \\
       50    &     1.51    &    1.32  &  0.12319 &  0.33276 &  3.862  &  12.682  \\
      100    &     1.52    &    1.33  &  0.12320 &  0.33251 &  3.862  &  12.705  \\
      150    &     1.55    &    1.33  &  0.12323 &  0.33213 &  3.864  &  12.737  \\
      200    &     1.57    &    1.34  &  0.12325 &  0.33169 &  3.866  &  12.773  \\
      250    &     1.58    &    1.36  &  0.12325 &  0.33130 &  3.867  &  12.802  \\
      300    &     1.62    &    1.36  &  0.12325 &  0.33081 &  3.870  &  12.838  \\ \hline\hline
   \end{tabular}
   \label{t:polarization}

\end{table*}

\begin{figure}
	\includegraphics[width = 9.0cm]{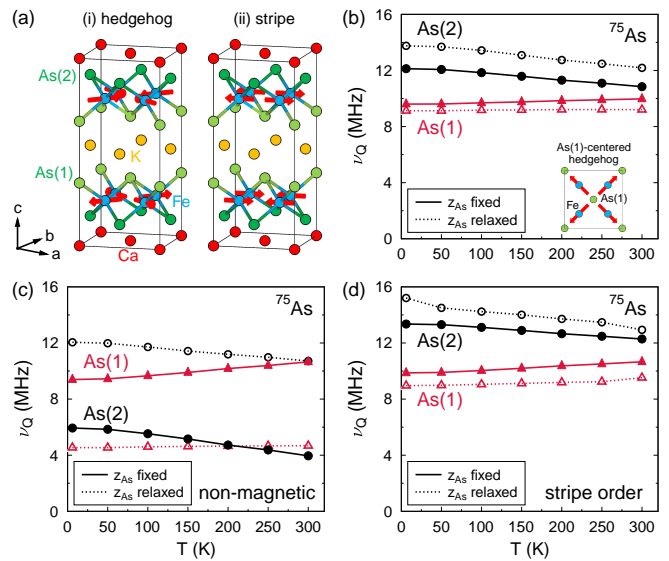}
	\caption{(a) Sketch of (i) the spin-vortex hedgehog and (ii) the stripe spin structure of \CaK. 
      Theoretical estimates of the As quadrupole frequencies $\nu_Q$ vs temperature (b) assuming an underlying spin-vortex hedgehog magnetic configuration (Fe spins in one of the FeAs layers are shown in the inset) compared to (c) the non-magnetic phase. 
     As(1) and As(2) stand for As sites near the K and Ca layers. Solid lines in (b) and (c) represent simulations with structures where the two As positions $z_\text{As}$ are fixed at the experimental values at room temperature, while the dashed lines correspond to the $z_\text{As}$ optimized in each case [with magnetism in (b) and without it in (c)]. (d) Calculated  $\nu_Q$  for the stripe order with fixed and optimized As positions.}
 \label{f:NMR_vs_temperature}
\end{figure}

    Calculations of $\nu_{\rm Q}$ in the hedgehog phase for both sets of structures show the same temperature trend as observed in our measurements [compare Fig.~\ref{f:NMR_vs_temperature}(b) with Fig.~\ref{fig:K}(b)]. 
    The first set of structures [$z_{As}$ fixed structures,  solid lines in Fig.~\ref{f:NMR_vs_temperature}(b)] indicate a significantly non-zero splitting between the two As frequencies at low temperature [$\Delta \nu_{\rm Q} = 2.5\:\text{MHz}$ at 6 K and $\nu_{\rm Q}(\mathrm{Ca}) > \nu_{\rm Q}(\mathrm{K})$],  while the splitting is almost zero in the room-temperature structure [$\nu_{\rm Q}(\mathrm{Ca}) \approx \nu_{\rm Q}(\mathrm{K})$]. 
    Our low-temperature estimate of  $ \Delta \nu_{\rm Q} $ is in a good agreement with the experimental result $\Delta \nu_{\rm Q} = 2.5\:\text{MHz}$,  although both calculated frequencies are $\sim$2.5 MHz lower than in experiment. 
     The second set of structures ($z_{As}$ relaxed structures, see Table~I)  show that relaxation  of the internal positions induces slight changes of the $z_\text{As}$ for As(1) and As(2) which  translates into a shortening of the As-As bonds across the Ca layer and their stretching across the K layer.
   Accordingly, in the optimized structures, the Fe magnetic moment is slightly diminished compared to the non-relaxed structures which is correlated with the overall smaller Fe-As distances (Table I). 
    As a function of temperature, the relaxed Fe-As distances barely change, which is in contrast to the  case of fixed $z_\text{As}$  where these distances decrease upon cooling.
    Upon structure optimization,  $\nu_{\rm Q}$ for the Ca layer is almost uniformly increased by $\sim$1.4 -- 1.8~MHz at all temperatures  [compare dashed and solid lines for As(2) in Fig.~\ref{f:NMR_vs_temperature}(b)], while $\nu_{\rm Q}$ for the K layer [As(1)] is much less affected. 
     The effect of $z_\text{As}$ optimization  on the NQR properties can be interpreted in terms of the Fe moments which change by only  $0.04\:\mu_\mathrm{B}$ with temperature, while they increase by $0.11\:\mu_\mathrm{B}$ and have, in general, a larger magnitude when the $z_\text{As}$ is fixed at all temperatures,  as highlighted in Table~I. Furthermore, the Fe-As and As-As bonding certainly affects the size  and ordering of $\nu_{\rm Q}$s, but it is not the only determining factor.

    Illustrative of the role of magnetic degrees of freedom is a comparison with the results for the truly non-magnetic state with zero Fe local moments [Fig.~\ref{f:NMR_vs_temperature}(c)]. 
    The order of the quadrupole frequencies is reversed leading to $\nu_{\rm Q}(\mathrm{Ca}) < \nu_{\rm Q}(\mathrm{K})$  and no crossing is observed at higher temperatures, which contradicts the experimental observations.  
    Optimizing $z_{As}$  for this non-magnetic state restores the correct order  $\nu_{\rm Q}(\mathrm{Ca}) > \nu_{\rm Q}(\mathrm{K})$ but the splitting $ \Delta \nu_{\rm Q} $ is severely  overestimated compared to the spin-polarized calculations and experiment.  
    In addition, one of the quadrupole frequencies is significantly underestimated by a factor  of two compared to the measured values of $\nu_{\rm Q}$. 
    Finally, from plot (d) in Fig.~\ref{f:NMR_vs_temperature}, we see that the stripe configuration provides similar values of the quadrupole frequencies as the hedgehog order. 
   The stripe order is close in energy to the hedgehog order, which is discussed in detail in Ref. \onlinecite{Meier2017}. 
   The quadrupole frequencies resulting from the magnetic fluctuations may be approximated by an average of the data shown in Figs.~\ref{f:NMR_vs_temperature} (b) and (d). 
   Such average would stay in good agreement with the measured $\nu_{\rm Q}$.

       These results show that both temperature-dependent structural features, non-zero Fe moment  and the type of magnetic fluctuations shape the temperature evolution of NQR in \CaK{}. 
    In this respect, different temperature behavior of the quadrupole frequencies for the two crystallographically inequivalent As sites might indicate the presence of $C_4$-type magnetic fluctuations competing with the usual stripe fluctuations, as suggested by the first-principles analysis. 
     Additional calculations (not shown here) suggest that the splitting between the quadrupole frequencies of As(1) and As(2) is sensitive to 
pressure, especially near the half-collapse transition around 4 GPa \cite{Kaluarachchi2017}.
   Finally it is noted that the origin of the opposite $T$ dependence of $\nu_{\rm Q}$ in CaFe$_2$As$_2$ and KFe$_2$As$_2$ is still an open question. 
   It is interesting to see that our approach based on the first-principles calculation explains this behavior, however disentangling the details of the various trends requires further investigations that are beyond the scope of the present study.



 \subsection{C. $^{75}$As spin-lattice relaxation rate 1/$T_1$}
 
\begin{figure}[tb]
  \includegraphics[width=8.0cm]{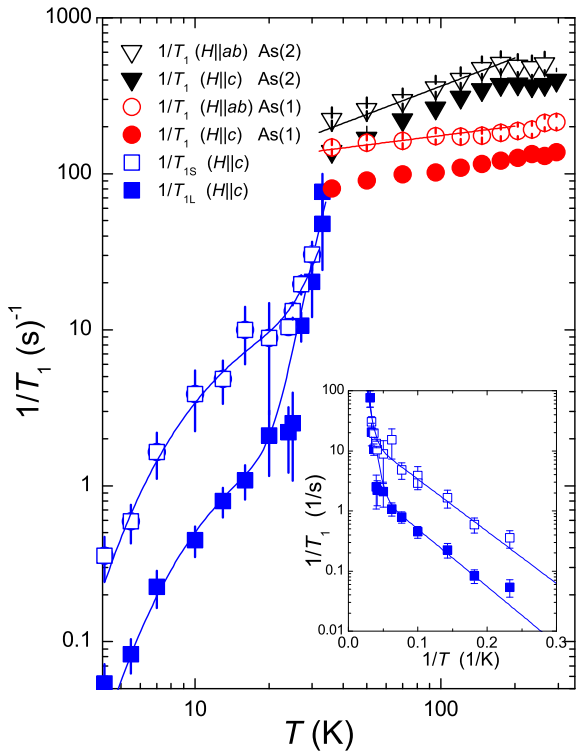} 
 \caption{Temperature  dependences of 1/$T_1$ of both As sites for $H$ $\parallel$ $c$ axis and $H$ $\parallel$ $ab$ plane. 
   The black and red lines show the power law temperature dependence with an exponent of 0.6 and 0.2  for the As(2) and As(1) sites, respectively. 
    The blue lines are calculated 1/$T_1$ with a two full gapped $s$ model using the SC gap parameters of $\Delta$ = 2.0 $\pm$ 0.4 meV and 20 $\pm$ 9 meV for $1/T_{1 \rm L}$ and $\Delta$ = 1.9 $\pm$ 0.4 meV and 16 $\pm$ 8 meV for $1/T_{1 \rm S}$.
    The inset shows the semi-log plots of $1/T_{1 \rm S}$ and $1/T_{1 \rm L}$ vs.  $1/T$ in the SC state.     
}
 \label{fig:1/T1}
  \end{figure}

        In order to  investigate magnetic fluctuations in CaK1144, we measured 1/$T_1$ at various temperatures.  
    Figure \ref{fig:1/T1} shows the $T$ dependences of 1/$T_1$ of both As sites for magnetic field directions, $H$ $\parallel$ $c$ axis and $H$ $\parallel$ $ab$ plane. 
    1/$T_1$ of the As(2) sites is nearly $T$ independent above $T^*$ $\sim$ 130~K for both $H$ directions.
     Below $T^*$, 1/$T_1$ decreases with a power law $T$ dependence of 1/$T_1$ $\propto$ $T^{0.6}$ as shown by the black line in Fig. \ref{fig:1/T1}.
     Similar $T$ dependence  of $^{75}$As 1/$T_1$ has been reported in $A$Fe$_2$As$_2$ ($A$ = K, Rb, Cs) where 1/$T_1$ increases with a power law 1/$T_1$ $\propto$ $T^{0.75}$ and levels off above $T^*$ $\sim$ 165~K, 125~K, and 85~K for A = K, Rb, and Cs, respectively \cite{Kondo2016}.
     Such a characteristic $T$ dependence of 1/$T_1$ is often observed in heavy fermion systems where the exponent strongly depends on the type of quantum criticality nearby, such as 1 for URu$_2$Si$_2$ \cite{Kohara1986}, 0.25 for CeCoIn$_5$ \cite{Kohori2001} and 0 for YbRh$_2$Si$_2$ \cite{Ishida2002}. 
     As for the As(1) sites, a similar $T$ dependence is also observed, but the exponent is found to be $\sim$ 0.2 which is smaller than 0.6 for the As(2) sites.
     Since the As(1) and As(2) are expected to probe the same spin dynamics from the Fe layers, one naively expects a similar $T$ dependence of 1/$T_1$ for both sites, although the magnitude can be different due to different hyperfine coupling constants.
   The results suggest that the Fe spins produce different magnetic fluctuations at  the As(1) and As(2) sites in CaK1144 and a possible origin of the different $T$ dependence of $1/T_1$ is discussed later.

      In the SC state below $T_{\rm c}$ $\sim$ 33~K, we attempted to measure 1/$T_1$ at the As(1) site for $H \parallel  c$ axis. 
      As described above, however, we could not distinguish the two As sites due to the broadening of the spectra. 
Therefore, $1/T_1$ is measured at the peak position of the spectrum at low $T$. 
    Because of the overlap of the spectra, we found two $T_1$ components: short ($T_{\rm 1S}$) and long ($T_{\rm 1L}$) relaxation times, which could be attributed to $T_1$ for the As(2) and As(1) sites, respectively.
    Just below $T_{\rm c}$, both 1/$T_{\rm 1L}$ and 1/$T_{\rm 1S}$  decrease rapidly without any coherence peak, consistent with an $s^{\pm}$ model.  
   With further decrease  of $T$,  both 1/$T_{\rm 1L}$ and 1/$T_{\rm 1S}$ show  broad humps around $\sim$ 15~K and exponential decreases at lower $T$, which can be clearly seen in the semi-log plot of 1/$T_1$ vs. 1/$T$ shown in the inset of Fig.~\ref{fig:1/T1}.   
   Similar behavior of 1/$T_1$ in SC state has been observed in LaFeAsO$_{1-x}$F$_x$ \cite{Oka2012} and K-doped BaFe$_2$As$_2$ superconductors \cite{Li2011, Matano2009}.
    These results suggest a multiple  fully-gapped SC state, consistent with previous measurements. 
    In fact, the $T$ dependence of 1/$T_1$ is roughly reproduced by a two full gap model with superconducting gap parameters $\Delta$ = 2.0 $\pm$ 0.4 meV and 20 $\pm$ 9 meV for 1/$T_{\rm 1L}$ and $\Delta$ = 1.9 $\pm$ 0.4 meV and 16 $\pm$ 8 meV for 1/$T_{\rm 1S}$   as shown by the solid lines. 
     The smaller gap $\sim$ 2.0  meV is not far from 2.5 meV obtained from the $\mu$SR measurements \cite{uSR2017} and in good agreement with 1-4 meV from the TDR and STM measurements \cite{Cho2017}.
     On the other hand, the larger gaps of 20 $\pm$ 9 and 16 $\pm$ 8 meV  seem to be greater than 6-9 meV from the other measurements \cite{Cho2017}.
    However, it should be noted that our estimates of the larger gap have a large uncertainty due to the poor statistics accuracy and relatively narrow $T$ range for the fit, therefore it should be regarded as tentative estimate.

\begin{figure}[tb]
 \includegraphics[width=8.6cm]{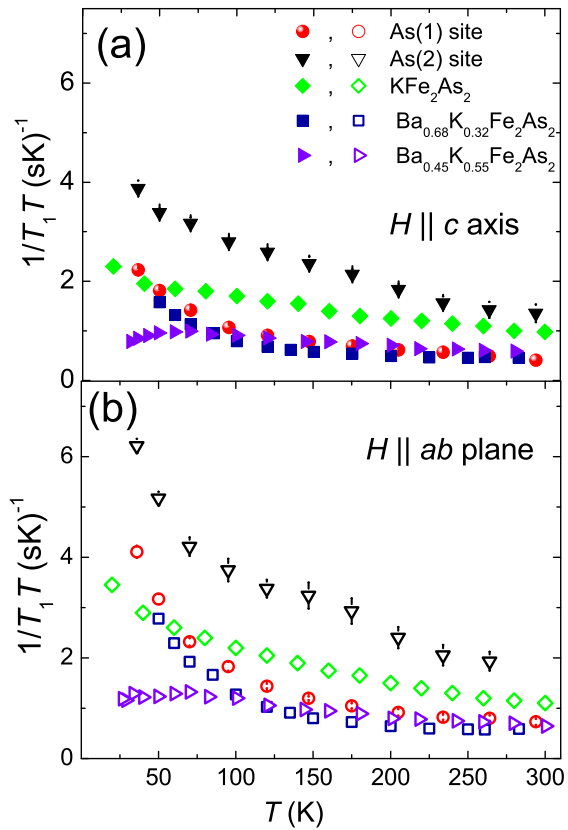} 
 \caption{(Color online)  Temperature dependence of 1/$T_1T$ in CaKFe$_4$As$_4$, together with those in Ba$_{1-x}$K$_x$Fe$_2$As$_2$ ($x$=0.32, 0.55, and 1).  (a) $H$ $\parallel$ $c$ axis. (b) $H$ $\parallel$ $ab$ plane.
}
 \label{fig:T1T}
 \end{figure}      

\begin{figure}[tb]
 \includegraphics[width=8.0cm]{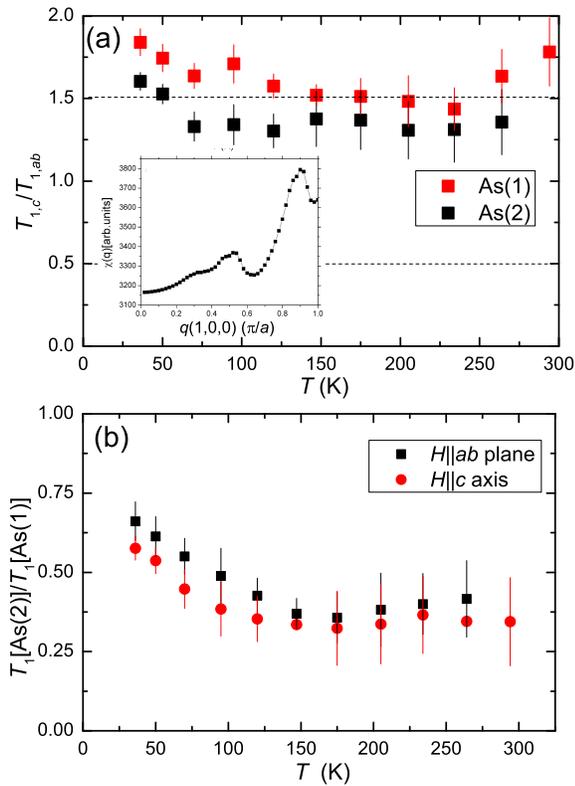} 
 \caption{(Color online)  (a) Temperature dependence of $r$ $\equiv$ $T_{1,c}$/$T_{1, ab}$ for the As(1) and As(2) sites. The horizontal dashed lines at $r$ = 1.5 and 0.5 correspond to the ratios for the AFM spin fluctuations with $q$ = ($\pi$, 0) [or (0, $\pi$)] and ($\pi$, $\pi$), respectively.  The inset shows the calculated $\chi(q)$ vs. $q$. (b) Temperature dependences of $T_1[{\rm As(2)]}$/$T_1[{\rm As(1)}]$ for$H$ $\parallel$ $c$ and $H$ $\parallel$ $ab$. 
}
 \label{fig:T1T-2}
 \end{figure}      

    Now we discuss magnetic fluctuations in the normal state.
    Figures \ref{fig:T1T} (a) and (b) show the $T$ dependences of $1/T_1T$ for $H$ $\parallel$ $c$ and $H$ $\parallel$ $ab$  in CaK1144. 
    For comparison, we also plot the $T_1$ data of  KFe$_2$As$_2$ ($T_{\rm c}$ = 3.5~K) \cite{Hirano2012},  Ba$_{0.45}$K$_{0.55}$Fe$_2$As$_2$  ($T_{\rm c}$ = 34~K) \cite{Hirano2012} and Ba$_{0.68}$K$_{0.32}$Fe$_2$As$_2$  ($T_{\rm c}$ = 38.5~K) \cite{Li2011}.
    1/$T_1T$ at the As(1) site is comparable to those of Ba$_{0.45}$K$_{0.55}$Fe$_2$As$_2$ and Ba$_{0.68}$K$_{0.32}$Fe$_2$As$_2$, while for the As(2) sites $1/T_1T$ is enhanced  larger than the KFe$_2$As$_2$ case. 
    All 1/$T_1T$ show a monotonic increase down to $T_{\rm c}$ with decreasing $T$ except for  Ba$_{0.45}$K$_{0.55}$Fe$_2$As$_2$ where 1/$T_1$ decreases at low $T$  showing gap-like behavior \cite{Hirano2012}.
     In general, 1/$T_1T$ is related to the dynamical magnetic susceptibility as $\frac{1}{T_1T}=\frac{2\gamma^{2}_{N}k_{\rm B}}{N_{\rm A}^{2}}\sum_{\vec{q}}|A(\vec{q})|^2\frac{\chi^{\prime\prime}(\vec{q}, \omega_0)}{\omega_0}$, where $A(\vec{q})$ is the wave-vector $\vec{q}$ dependent form factor and $\chi^{\prime\prime}(\vec{q}, \omega_0)$ is the imaginary part of the dynamic susceptibility at the Larmor frequency $\omega_0$. 
     Although the exponents of the power-law behavior of 1/$T_1$ for As(1) and As(2) are different, we can  attribute the increase of 1/$T_1T$ to the growth of AFM spin fluctuations at $q\neq0$  since both $K_c$ and $K_{ab}$, reflecting $\chi^{\prime}(q=0, \omega_0=0)$, are nearly independent of $T$.

   In order to see the nature of spin correlations, we plotted the ratio of 1/$T_1$ for the two field directions, $r$ $\equiv$ $T_{1,c}$/$T_{1, ab}$. 
  According to previous NMR studies performed on Fe pnictides and related materials \cite{KitagawaSrFe2As2, KitagawaAFSF, Hirano2012, PandeySrCo2As2}  
   the  ratio $r$  depends on AFM spin correlation modes as  
 \begin{eqnarray}
 r =  \left\{
 \begin {array}{ll}

 0.5 + \left(\frac{S_{ab}}{S_c}\right) ^2    \mbox{~ for}  ~ {\bf q} = (\pi,0)  \mbox{~or} ~ (0, \pi) \label{eqn:stripcorrelation}\\

 0.5     \mbox{~ for}   ~ {\bf q} = (\pi,\pi) \label{eqn:Neelcorrelation}\\
 \end {array}
 \right .
 \label{eqn:correlation}
 \end{eqnarray}
where  ${\cal S}_{\alpha}$ is the amplitude of the spin fluctuation spectral density at the NMR frequency along the $\alpha$ direction. 
The wavevectors are given in the single-iron Brillouin zone notation. 

    As plotted in Fig.\ \ref{fig:T1T-2}(a), the ratios for both the As(1) and As(2) sites are close to $r$ $\sim$ 1.5.
    This indicates that the spin fluctuations in CaKFe$_4$As$_4$ are characterized by the AFM spin fluctuations with  $q$ = ($\pi$, 0) or (0, $\pi$) and ${\cal S}_{c}$ $\sim$ ${\cal S}_{ab}$ indicating  the isotropic spin fluctuations.
    In order to confirm the AFM spin correlations, we have calculated $\chi(q)$ \cite{Yongbin}  using the full-potential linearized augmented plane wave method \cite{Blaha2001} with a generalized gradient approximation \cite{Perdew1996}.
    The calculated $\chi(q)$ is shown in the inset of Fig.\ \ref{fig:T1T-2}(a), which actually exhibits the AFM spin correlations where a peak around  the wavevector ${\bf q}$ = ($\pi$, 0) or (0, $\pi$)  is observed with a small structure.
     A similar structure in $\chi(q)$ was reported in CaFe$_2$As$_2$ \cite{Goldman2009}. 

     In most of the Fe pnictide SCs, the AFM fluctuations are described in terms of the stripe-type spin fluctuations.
       Recently another possible AFM fluctuation with  $q$ vectors [($\pi$, 0) and (0, $\pi$)] has been proposed in the spin vortex crystal  state where Fe spins are non-collinear and form spin vortices staggered across the iron square lattices \cite{Fernandes2016}.
     Since the stripe- and spin vortex-type states produce the AFM spin correlations with the same $q$, in general, it would be difficult to distinguish between the two different spin correlations from the $T_1$ measurements. 
    However, \CaK   provides us a special occasion to distinguish between them.
   In the hedgehog vortex ordered state, the internal fields at As(1) is calculated to be finite along the $c$ axis while the internal field at As(2) is zero  due to a cancellation originating from the characteristic spin structure \cite{Meier2017}.  
     Therefore, one expects that 1/$T_1T$ for As(1) is more enhanced than that for As(2) if the AFM spin fluctuations originate from the spin vortex hedgehog-type spin correlations.  
   On the other hand, in the case of stripe-type AFM fluctuations, the temperature dependence of $1/T_1T$ for As(1) should scale to that of $1/T_1T$ for As(2) sites since there is no cancellation of the internal field at both As sites  \cite{Meier2017}. 
     1/$T_1T$ for As(1) divided by 1/$T_1T$ for As(2) for $H$ $\parallel$ $c$ and $H$ $\parallel$ $ab$ are shown in Fig.\ \ref{fig:T1T-2}(b) where clear enhancements are observed below $\sim$ 150 K. 
  Above $\sim$ 150 K, the ratios of $T_1[{\rm As(2)}]/T_1[{\rm As(1)}]$  show a nearly temperature independent value of $\sim$ 0.35 which could be determined by the different hyperfine coupling constants for the As(1) and As(2) sites. 
    The increases below $\sim$ 150 K indicate  that the As(1) sites pick up the AFM  spin fluctuations more strongly than the As(2) sites, consistent with the case of hedgehog-type spin fluctuations.
    Our {\it ab initio} DFT total energy calculations  show that both stripe order and spin-vortex order are energetically competing in CaKFe$_4$As$_4$ and the temperature dependence of $\nu_{\rm Q}$, as presented above, is well described by assuming a hedgehog spin vortex order. 
    Furthermore,  a hedgehog spin vortex order has been recently observed in Co-doped and Ni-doped CaKFe$_4$As$_4$~\cite{Meier2017}.
     Therefore, our results strongly indicate the existence of the spin vortex hedgehog-type spin correlations in the paramagnetic state. 
      Further experimental and theoretical studies are important to elucidate more details of the characteristic of the hedgehog-type spin correlations and also its relationship to superconductivity in CaKFe$_4$As$_4$ and the carrier-doped CaKFe$_4$As$_4$.

 \section{IV. Summary}  

     In conclusion, we have carried out $^{75}$As NMR measurements in the single crystalline CaKFe$_4$As$_4$. 
    The observation of two distinct $^{75}$As NMR lines indicates that K and Ca ions are not randomly  distributed but are well ordered forming the Ca and K layers without a significant site inversion. 
      We found that $\nu_{\rm Q}$ of the As(1) and As(2) sites show an opposite $T$ dependence. 
      Based on our \textit{ab initio} density functional theory calculations, this behavior might originate from the presence 
      of spin vortex hedgehog-type spin fluctuations which represent a superposition of ($\pi$, 0) and (0, $\pi$) 
      stripe fluctuations. We find that the calculated quadrupole frequencies are comparably affected 
      by the Fe-As and As-As bonding, non-zero Fe moments and the underlying magnetic order which 
      simulates the spin fluctuations to a first approximation. 
  In the normal state, we also observe evidence  of the enhancement of isotropic AFM spin fluctuations with  wavevector ${\bf q}$ = ($\pi$, 0) or (0, $\pi$) in  
       the $T$ dependences of 1/$T_1T$ and $K$. 
      It is suggested that the different $T$ dependence of 1/$T_1$ for the As(1) and As(2) sites is explained by the peculiar spin fluctuations due to the
spin vortex hedgehog-type spin correlations.
   In the SC state, we observed    a sudden decrease in $K$, indicating spin-singlet Cooper pairs. 1/$T_1$ shows a rapid decrease 
      below $T_{\rm c}$ without showing  a Hebel-Slichter peak and decreases exponentially at low 
      temperatures. These results indicate an $s^{\pm}$ nodeless two-gap SC state.

 \section{V. Acknowledgments}
The authors would like to thank Khusboo Rana for his assistance with NQR experiments. 
We also thank Paul Wiecki, Rafael Fernandes, Peter Orth, Igor Mazin and Andreas Kreyssig for helpful discussions. 
The research was supported by the U.S. Department of Energy, Office of Basic Energy Sciences, Division of Materials Sciences and Engineering. Ames Laboratory is operated for the U.S. Department of Energy by Iowa State University under Contract No.~DE-AC02-07CH11358. 
WRM was supported by the Gordon and Betty Moore Foundations EPiQS Initiative through Grant GBMF4411. VB and RV
are supported by the Deutsche Forschungsgemeinschaft
through grant SFB/TRR 49 and the computer time was allotted by the centre for
supercomputing (CSC) in Frankfurt. 

\bigskip*\ present address: Department of Chemistry, Princeton University, Princeton, NJ 08544, USA

\end{document}